\documentclass[10pt]{article}
\usepackage[dvips]{graphics}
\begin{document}

\title{Hadronic Molecules in Lattice QCD}
\author{Chris Stewart and Roman Koniuk \\
Department of Physics \& Astronomy \\ York University, 4700 Keele 
St., Toronto, Ontario M3J 1P3, Canada.}     
\date{\today}

\maketitle

\begin{abstract}
An adiabatic approximation is used to derive the binding potential 
between two heavy-light mesons in quenched SU(2)-colour lattice QCD. 
Analysis of the meson-meson system shows that the potential is attractive at 
short- and medium-range. 
The numerical data is consistent with the Yukawa model of pion exchange. 
\end{abstract}

\newcommand{\be}{\begin{equation}}
\newcommand{\ee}{\end{equation}}
\newcommand{\bea}{\begin{eqnarray}}
\newcommand{\eea}{\end{eqnarray}}
\newcommand{\Tr}{\mbox{Tr}}
\newcommand{\bc}{\begin{center}}
\newcommand{\ec}{\end{center}}
\newcommand{\x}{\vec{x}}
\newcommand{\R}{\vec{R}}

\unitlength=1.0cm

\section{Introduction}
Quantum Chromodynamics is widely accepted as the correct theory of 
quark-quark interactions.
QCD effective theories such as Chiral Perturbation Theory have successfully 
described a wide range of hadronic phenomena.
Nucleon-nucleon interactions, for example, have traditionally been modeled 
by effective meson-exchange theories. Such theories are long-range 
approximations to QCD interactions mediated by gluons 
and quark-pairs. A derivation of the strong nuclear force from full QCD 
is obviously desirable.

Lattice QCD provides a first-principles framework for such a derivation. 
However, the nucleon-nucleon system, containing six light quarks, is well 
beyond the current limits of computation. 
In fact, at first sight such a computation seems impossible---Lattice QCD is 
providing physical hadronic data consistent with experiment at the 
$\sim 5 $\% level. The nuclear physics problem, the interaction between 
two colour-neutral objects, is an effect two or three orders of magnitude 
smaller. Nevertheless, nuclear models involving
static quark clusters in unquenched QCD \cite{clust} and light-quark hadrons in
confining QED$_{2+1}$ \cite{QED} have been examined, and both give
encouraging results. An unquenched lattice QCD simulation of the 
nucleon-nucleon system is doubtless some years away. In this paper we 
examine the same fundamental problem in a simpler setting, by examining the 
interactions between two heavy-light mesons \cite{Fiebig}.   

An elementary problem in molecular physics is the derivation of the 
binding energy of the $\mbox{H}_2$ molecule. A good first approximation 
can be found by using an adiabatic approximation, in which the (slow) nuclear 
kinematics
are removed by treating each nucleus as a static force centre, 
and the remaining (fast) electronic problem is solved as a function of the 
nuclear separation. This generates an effective potential for the two
nuclei, which can be used as input for the internucleon 
wavefunctions and energy levels.

We wish to consider an analogous problem---a `hadronic molecule' consisting of
two heavy-light mesons, in which the heavy quarks are treated as static
colour sources, playing the role of the (slow) atomic nuclei. 
The gluons and light-quarks play the role of the fast degrees of freedom. 
In addition to the static heavy-quark gluon-exchange interaction, this 
calculation will include the effects of interactions between gluons and 
light-quarks, as well as light-quark exchange. 
The total energy of the system
can be found from the asymptotic long-time behaviour of the meson-pair
propagator. An effective potential is extracted as a function of the meson 
separation. 

The system described above contains only two light valence quarks. As such, 
we are still quite removed from a direct simulation of the nucleon problem. 
However, as all nuclei are hadronic molecules, our qualitative conclusions 
should be somewhat universal. Simulations of $MM$- and $M\bar{M}$-systems 
are of further interest 
because of their immediate application to $K\bar{K}$ phenomena. 
Studies of multi-quark states indicate that the only likely bound 
four-quark systems are mesonic molecular states \cite{4quark}.
Two exotic particles, the
$a_0(980)$ and $f_0(975)$, are thought to be lightly-bound $K \bar{K}$ 
molecules \cite{kkbar}. 

This paper describes a lattice simulation of heavy-light mesonic molecules, 
using SU(2)-colour quenched QCD for computational simplicity. We consider both
meson-meson and meson-antimeson systems.
The next section describes the 
meson operators used in this simulation, and outlines the various
contributions to the meson-pair propagators. 
In Section III we list the Lattice actions used and the values of simulation 
parameters, and our results are given in Section IV.
 
\section{Operators and propagators}
We used a local pseudoscalar meson operator,
\be
M(\x,t) = \bar{\psi}_{l}(\x,t) \gamma_{5} \psi_{h}(\x,t) \, ,
\ee
constructed from heavy ($h$) and light ($l$) Wilson quarks. 
With quark propagators given by
\be
G(\x,t;0,0) = \langle 0 | \psi(\x,t) \bar{\psi}(0,0) | 0 \rangle \,,
\ee
the zero-momentum meson propagator is then
\bea
\label{mesprop}
G_{M}(t) &=& \Tr \sum_{\x} G_{h}(\x,t;0,0) \, \gamma_{5} \, G_{l}(0,0;\x,t) 
\gamma_{5} \nonumber \\
 &=& \Tr \sum_{\x} G_{h}(\x,t;0,0) \, G_{l}^{\dag}(\x,t;0,0) \, ,
\eea

The hopping-parameter expansion of the quark propagator \cite{Rothe} gives, 
in the limit of infinite quark mass,
\be
\label{heavyq}
G_{h}(\x,t;0,0) \propto \delta^3_{\x,0} \prod_{\tau = 0}^{t-1} U_{4}(0,\tau) \, .
\ee
That is, the heavy-quark propagator becomes simply a string of gauge-field
links in the time direction, and is calculated in one sweep of the lattice. 
Figure 1 shows a contribution to the heavy-light meson propagator.

An operator for a heavy-light meson-pair with spatial separation $\vec{R}$ is
\be
{\cal O}_{MM}(\vec{R},t) = M(\x,t)\, M(\x+\vec{R},t) \,.
\ee
The propagator for this system is then
\be
\label{mmprop}
G_{MM}(t,\vec{R}) = G_{D}  - G_{E} \, ,
\ee 
where a sum over all contractions of the quark fields was performed, giving
\bea
G_{D}(t,\R) &=& \Tr \left [ G_{h}(0,t;0,0) \, G^{\dag}_{l}(0,t;0,0) 
\right ] \nonumber \\
& & \times \Tr \left [ G_{h}(\R,t;\R,0) \, G^{\dag}_{l}(\R,t;\R,0) \right ] 
\nonumber \,, \\
G_{E}(t,\R) &=& \Tr \left [ G_{h}(0,t;0,0) \, G^{\dag}_{l}(\R,t;0,0) 
\right . \nonumber \\
& & \left . \, G_{h}(\R,t;\R,0) \, G^{\dag}_{l}(0,t;\R,0) \right ] \, .
\eea
Note that, with the heavy-quark approximation (\ref{heavyq}), 
the separation $\R$ is 
well-defined throughout the propagation.
Contributions to $G_{D}$ and $G_{E}$, the direct and exchange terms, 
are depicted in Figure 2.

The meson-antimeson propagator has a similar form,
\be
\label{mmbarp}
G_{M\bar{M}}(t,\R) = {\cal G}_{D} - {\cal G}_{E} \, ,
\ee 
where
\bea
{\cal G}_{D}(t,\R) &=& \Tr \left [ G_{h}(0,t;0,0) \, G^{\dag}_{l}(0,t;0,0) 
\right ] \nonumber \\
& & \times \Tr \left [ G_{l}(\R,t;\R,0) \, G^{\dag}_{h}(\R,t;\R,0) \right ] 
\nonumber \,, \\
{\cal G}_{E}(t,\R) &=& \Tr \left [ G_{h}(0,t;0,0) \, 
G^{\dag}_{l}(\R,t;0,0) \right .
\nonumber \\
& & \left .G^{\dag}_{h}(\R,t;\R,0) \, G_{l}(0,t;\R,0) \right ] \, .
\eea
Contributions to these terms are shown in Figure 3.

The ground-state energy of the two-meson system is extracted from 
the long-time evolution of the system's propagator \cite{Rothe}. 
Expanding the propagator
as a sum over eigenstates of the Hamiltonian $|N\rangle$ with eigenvalues 
$E_N$, the propagator becomes proportional to the exponential of the 
ground-state energy $E_0$ for long times,
\bea
G(R,t) &=& \langle 0 | \bar{{\cal O}}_{MM}(\R,t) {\cal O}_{MM}(\R,0)
|0 \rangle \nonumber \\
&=& \sum_{N} \langle 0 | \bar{{\cal O}}_{MM}(\R,t) | N \rangle 
\langle N | {\cal O}_{MM}(\R,0) | 0 \rangle \nonumber \\
&=& \sum_{N} \left | \langle 0 | \bar{{\cal O}}_{MM}(\R,0) | N \rangle 
\right |^2 e^{-E_{N}(R)t} \nonumber \\
&\simeq& c_0 \, e^{-E_{0}(R)t} \, \mbox{ as } \, t \rightarrow \infty \, .
\eea
The binding energy $V(R)$ is then defined as the difference between the 
ground-state energy and the masses of the two independent mesons,
\be
E_{0}(R) = 2m_{M} + V(R) \,.
\ee

\section{Simulations} 
A model of quenched QCD with $SU(2)$-colour was used to simplify analysis.
The standard Wilson action for the pure gauge field is
\cite{Rothe},
\be
S_{G} = -\beta \sum_{P} \left ( 1-\frac{1}{2} \Tr \, U_{P} \right ) \, .
\ee
Here $\beta$ is the inverse of the square of the coupling strength, and the 
sum is over all plaquettes on the lattice.
The Wilson action for light quarks has the form
\be
S_{q} = \sum_{n,m} \bar{\psi}_{n}\, K_{nm}\, \psi_{m} \, ,
\ee
with the action matrix given by
\bea
K_{nm} &=& \delta_{n,m} - \kappa \sum_{\mu} \Big [ (1-\gamma_{\mu}) 
\, U_{\mu}(n) \, \delta_{n+\hat{\mu},m} \nonumber \\
&  & + \, (1+\gamma_{\mu}) \, U^{\dag}_{\mu}(n-\hat{\mu}) \,  
\delta_{n-\hat{\mu},m}) 
\Big ] \, .
\eea
The light-quark propagator is then the inverse of the action matrix,
\be
\label{lprop}
G_l(n,m) = K^{-1}_{nm}
\ee

Simulations were performed on an $8^4$ lattice with periodic boundary 
conditions. An ensemble of 400 gauge field configurations was generated 
with a Metropolis algorithm at $\beta = 2.3$. Using the standard
prescription for translating physical SU(3) QCD to unphysical SU(2) QCD, we
used the pion and rho meson masses to set the lattice scale \cite{Wol}. 
We found the lattice spacing to be roughly $0.2$ fm, and the lattice is 
then $\sim 1.6$ 
fm along a side. One expects the radius of the heavy-light meson to be
around half that 
of the pion, or $\sim 0.5$ fm, and so our
lattice volume should be large enough to contain the two mesons.
 
To ensure statistical
independence, 200 update sweeps were performed between stored configurations.
Quark propagator matrix inversions (\ref{lprop}) were performed with a 
conjugate gradient algorithm, at a value of $\kappa = 0.165$ for the hopping 
parameter.

\section{Results}
All propagators under consideration are symmetric in time, since the 
periodic boundary conditions allow contributions from quarks and antiquarks 
winding around the lattice. The mass of
the heavy-light meson was taken from fits to the meson propagator
(\ref{mesprop}),
\be
G_M (t) = c_M \, \cosh\left( m_M \left(t-T/2 \right) \right) 
\, ,
\ee
where $T=8$ is the period of the lattice in the time direction. While this is
an asymptotic relation, valid for long times, in practice fits with 
acceptable $\chi^2$ were obtained for 
$2 \leq t \leq 6$. The average mass obtained from these fits
was 
\be
\label{mesmass}
m_M = 0.969 \pm 0.008
\ee
in lattice units.

\subsection{The $MM$-system}
The meson-pair propagator, (\ref{mmprop}), was constructed for values of the 
separation $R$ from $0$ to $7$ lattice spacings. 
Given that statistical fluctuations in the individual
meson propagators will likely be highly correlated to fluctuations in the 
$MM$-system propagator, we were able to reduce statistical errors by analysing
the ratio
\be
\label{rat}
C(\R,t) = \frac{G_{MM}(\R,t)}{G_{M_1} \, G_{M_2}} \,.
\ee

For each value of the separation $R$, the configuration average of $C(\R,t)$ 
was fit with a ratio of cosh functions,
\be
C(R,t) = c_R \, \frac{\cosh^2 \left((2m_M - V(R)\right) \, \tau)}
{\cosh^2 (m_M \, \tau)} \, , 
\ee
with
\be
\tau = t-\frac{T}{2} \, ,
\ee
and using the meson mass given in (\ref{mesmass}). 
The results of the fit for 
$R = 0$ and $R=1$ are shown in Figure 4. 

The binding potential extracted from (\ref{rat}) is shown in Figure 5.
The data has been fit with a periodic Yukawa function,
\be
\label{yuk}
V(R) = a\, \left( \frac{e^{-\mu r}\left(1-e^{-r/d}\right)}{r} + 
\frac{e^{-\mu (N-r)}\left(1-e^{-(N-r)/d}\right)}{N-r}\right) \, ,
\ee
where $N$ is the spatial length of the lattice.
The parameter $d = d(N)$ provides the correct limiting behaviour for
$r = 0$. This form 
reduces to the continuum Yukawa function as $N \rightarrow \infty$ 
($d \rightarrow 0$), and is in excellent agreement
with numerical calculations of the discrete Fourier transform  
of $\frac{1}{k^2 + \mu^2}$. We found $d \simeq 0.5$ at $N=8$, and using this 
in (\ref{yuk}) we obtain (goodness-of-fit $\chi^2/\mbox{(degree of freedom)}
 = 0.5$)
\bea
\label{yukpars}
a &=& (-12 \pm 3) \times 10^{-2} \,m_M \, ,\nonumber \\
\mu &=& 0.7 \pm 0.4 \, \mbox{ (lattice units)}.
\eea
The strength of this potential, described by the constant $a$, is consistent 
with nuclear potentials which are two orders of magnitude smaller than nucleon 
rest-masses.

The potential is certainly attractive for small separations.
Evidence for attraction at small separations was also seen in the 
staggered-quark calculation of Mih\'{a}ly \emph{et.~al.} \cite{Fiebig}.
To ascertain whether the potential is significantly non-zero for meson 
spacings $R > 0$, the data for $R = 2$ to $R = 6$ was fit with a constant 
function,
\be
V(R) = c \, ,
\ee
giving
\be
c = (-1.3 \pm 0.8) \times 10^{-2} \,m_M \,.
\ee
This constant term is more than one and a half standard deviations below zero,
suggesting that there is some attractive potential even at medium-range 
separations. 
We regard the near-vanishing potential at the largest 
separations ($R=4$) to be circumstantial evidence that our lattice volume is 
large enough to contain the meson-meson system.

The Yukawa parameters (\ref{yukpars}) correspond to a model of hadron 
interactions mediated by 
pion-exchange, with the exchanged pion mass given 
by $m_{\pi} = \mu = 0.7 \pm 0.4$ in lattice units. Even though we are working
in the quenched approximation and virtual $q\bar{q}$-pairs are absent, 
the exchange term in (\ref{mmprop})
will include pion exchange contributions---one such exchange is shown in 
Figure 6. 

Pion propagators were constructed using the same gauge field configurations
at the same value of $\kappa$, to provide a pion mass for direct 
comparison with (\ref{yukpars}). The pion propagator is equivalent 
to (\ref{mesprop}), with two light-quark fields,
\be
G_{\pi}(t) = \Tr \sum_n G^2_l(n,t;0,0) \, ,
\ee
and the pion mass was extracted from cosh fits to this propagator.
The result of $m_{\pi} = 0.38 \pm 0.18$ agrees within 
errors with the Yukawa pion mass above. 

\subsection{The $M\bar{M}$-system}
Simulations of a meson-antimeson system using (\ref{mmbarp}) were performed. 
The results were not consistent with those described above for the $MM$-system.
The propagators tended to behave exactly like Wilson Loops, just indicating 
that the system's ground state was a state where the light quarks have 
annihilated, leaving a static $Q\bar{Q}$ pair. Figure 3(b) shows a 
contribution from such a state. 
A proper simulation of $M\bar{M}$ systems may require a more realistic,
dynamic treatment of the heavy quarks. 

\section{Conclusions}
The binding potential between two heavy-light mesons in SU(2) QCD was found 
by considering the long-time behaviour of propagators for the two-meson system.
The heavy-quark approximation simplified analysis considerably, as the heavy 
quark plays a role analogous to the 
heavy nucleus in the adiabatic calculation of the binding potential in the 
$\mbox{H}_2$ molecule.

The binding potential derived in this study is attractive and short ranged, 
and periodic due to the boundary conditions of the lattice. 
The meson-meson interaction can be modeled by a pion-exchange Yukawa theory.
Comparing our results with a Yukawa model, the potential has the 
correct form, and the exchanged pion mass derived from the potential agrees
within errors 
with the pion mass calculated at the same values for the simulation 
parameters. While the large errors prompt caution, these results make
us confident that the heavy-light meson-pair system is tractable in lattice
QCD, and is a promising step towards a first principles treatment of 
the nuclear physics problem.   

Unfortunately, in the $M\bar{M}$-propagator we only saw evidence of a static 
$Q\bar{Q}$-pair, due to the annihilation of the light-quarks. Overlap with 
physical $M\bar{M}$-states would be enhanced if the heavy-quarks were allowed 
to propagate spatially,
perhaps in the regime of non-relativistic QCD. More work needs to be done 
to fully appreciate the subtleties of this system. 

The very low light-quark mass used in this study 
($\kappa=0.165$ gives a 
rho-pion mass ratio of $\frac{m_\rho}{m_\pi} \simeq 3.0$) 
resulted in large statistical fluctuations in the quark propagators,
evidenced by the error bars in
Figure 5. Unfortunately, raising the quark mass leads to smaller
contributions from light-quark exchange, and so weakens the resulting 
potential. 
These problems are in part due to the low-order approximations 
in the Wilson quark and gluon actions. We have
commenced simulations using improved actions for the gauge fields
\cite{imp1,imp2} and fermion fields \cite{imp3}. 
These actions remove most of the discretisation
errors, and restore much of the rotational symmetry of the continuum
QCD action lost in the hypercubic lattice approximation.

\newpage

\begin{figure}
\begin{center}
\scalebox{0.8}[0.8]{\includegraphics{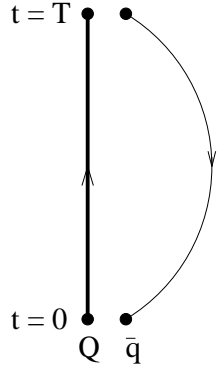}}
\caption{Heavy-light meson propagator constructed from heavy ($Q$) and light
(q) quark propagators.}
\end{center}
\end{figure}

\begin{figure}
\begin{center}
\scalebox{0.8}[0.8]{\includegraphics{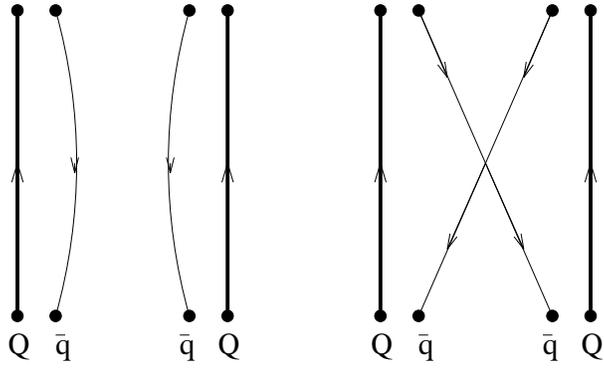}}
\caption{(a) Direct and (b) Exchange terms for $MM$-system.}
\end{center}
\end{figure}

\begin{figure}
\begin{center}
\scalebox{0.8}[0.8]{\includegraphics{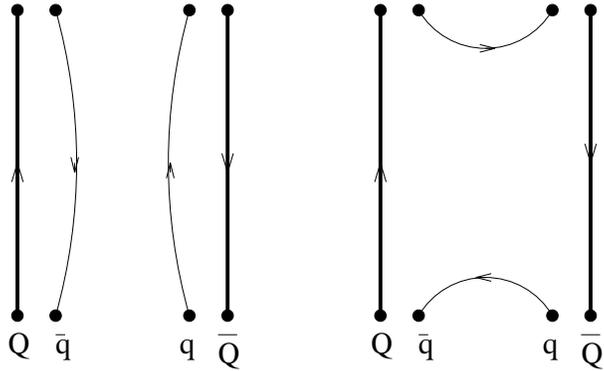}}
\caption{(a) Direct and (b) Exchange terms for $M\bar{M}$-system.}
\end{center}
\end{figure}

\begin{figure}
\begin{center}
\scalebox{0.7}[0.7]{\includegraphics{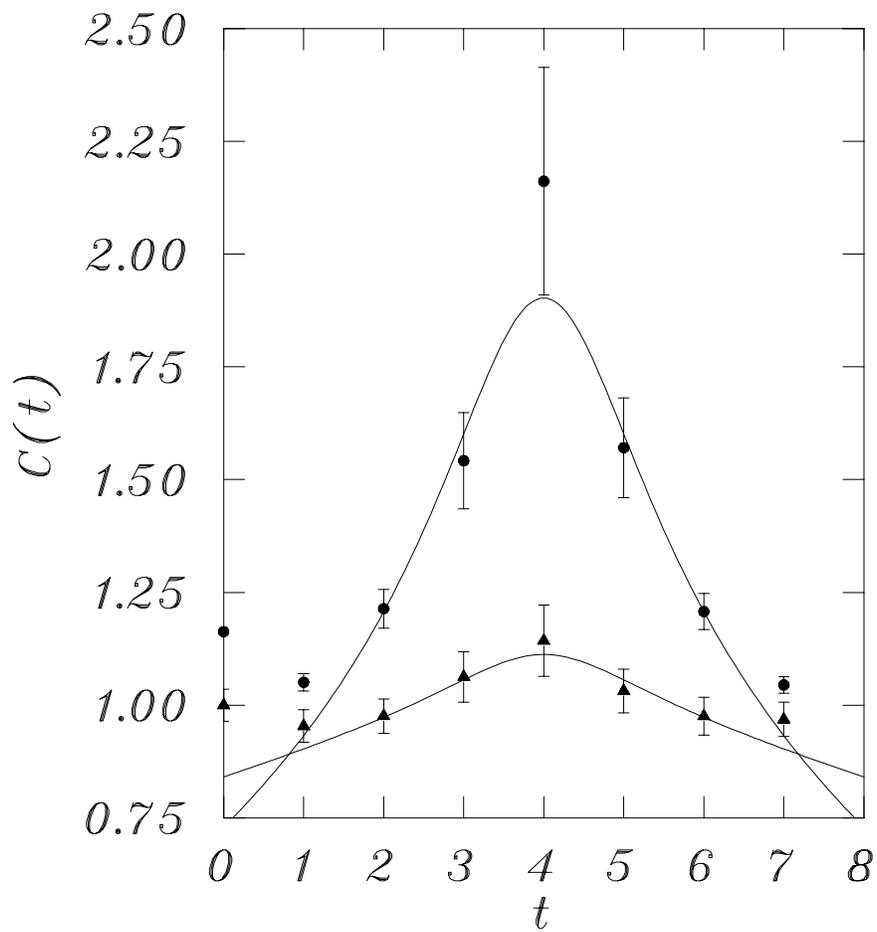}}
\caption{Fits to the propagator ratio (19), 
for meson separation $R=0$ (circles) and $R=1$ (triangles).
Error bars indicate Jacknife errors.}
\end{center}
\end{figure}

\begin{figure}
\begin{center}
\scalebox{0.7}[0.7]{\includegraphics{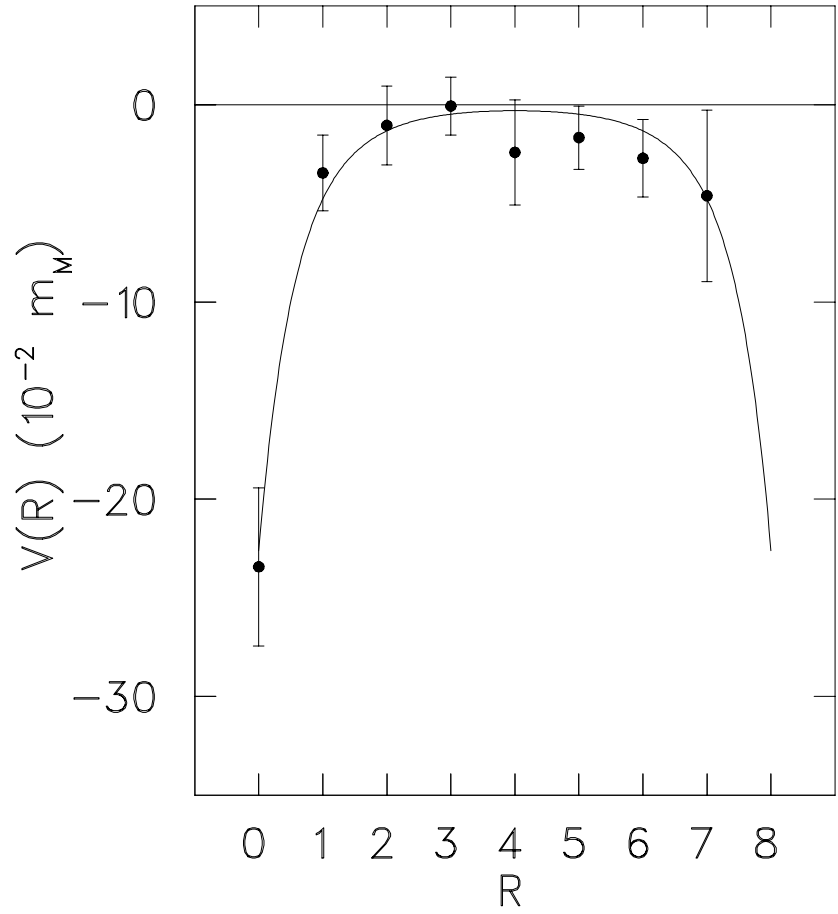}}
\caption{$MM$ potential $V(R)$ fit with a discrete Yukawa function (21).}
\end{center}
\end{figure}

\begin{figure}
\begin{center}
\includegraphics{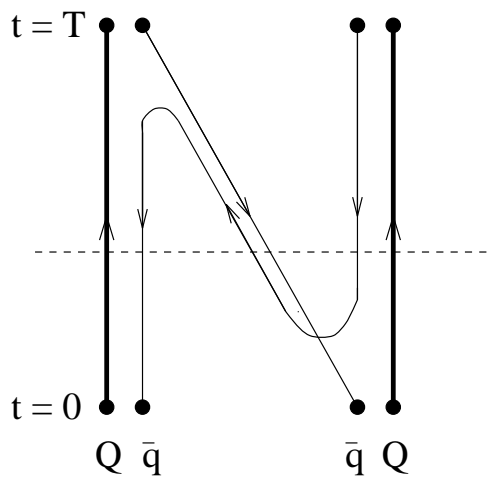}
\caption{A pion-exchange contribution to the $MM$-propagator---at the 
time-slice indicated by the dotted line, two heavy-light mesons and 
a pion are present.}
\end{center}
\end{figure}

\end{document}